\documentclass[sigconf]{acmart}
\pdfoutput=1
\usepackage{booktabs} 

\usepackage{algorithm}  
\usepackage{algpseudocode}  
\usepackage{amsmath}  

\usepackage{diagbox}
\begin{document}
\title{Congestion Control for RTP Media: a Comparison on Simulated Environment}
\titlenote{Produces the permission block, and
  copyright information}

\author{Songyang Zhang}
\affiliation{
  \institution{School of Computer Science and Engineering, Northeastern University, China}
}
\email{sonyang.chang@foxmail.com}
\author{Weimin Lei}
\affiliation{
  \institution{School of Computer Science and Engineering, Northeastern University, China}
}
\email{leiweimin@ise.neu.edu.cn}
\begin{abstract}
To develop low latency congestion control algorithm for real time taffic has been gained attention recently. RTP Media Congestion Avoidance Techniques (RMCAT) working group was initiated for standard defination. There are three algorithms under this group, Network Assisted Dynamic Adaptation (NADA) proposed
by Cisco, Google Congestion Control (GCC) proposed
by Google and Self-Clocked Rate Adaptation for Multimedia(SCReAM) proposed by Ericsson. This paper compares and analyses the performance of these algorithms on simulation environment. Results show GCC has well fairness property and performs well in lossy link but slow convergence in dynamic link, NADA stabilizes its rate quickly but suffers "late-comer" effect, SCReAM has the lowest queue occupation but also lower link capacity utilization.
\end{abstract}

\keywords{congeation control, real time traffic, ns3 simulation}

\maketitle

\section{Introduction}

Pioneered by Jocobson’s work\cite{jacobson1988congestion}, which later developed into TCP Reno algorithm, 
network congestion control has been an unfading topic in computer networks research. 
The control law proposed by Jocobson is to regulate TCP sending rate according to additive increase and multiplicative decrease(AIMD). On every RTT, the sender would send one more packet into network and multiplicatively reduces congestion window size by half when packet loss happens.
From then on, most of the research works such as Bic\cite{xu2004binary}, Cubic\cite{ha2008cubic}, were mainly focused on TCP improvement or adapted the basic AIMD control law to different network environment.

The congestion control algorithms in TCP are mainly compliant with bulk data transfer. TCP saw-tooth sending rate pattern caused by congestion window decrease and its retransmission to guarantee reliablity introducing large end to end delay make it unfit for time stringent traffic transmission. The real time video taffic is quite sensitive  to connection latency but can suffer some packets loss to some extent, so RTP-based media usually streams over UDP and implements its congestion control mechanism on application layer.

Today networks are used in a different manner, most flows on the network are real-time, delay sensitive traffic such as video conference, web-browsing, gaming. Video currently accounts for 70\% of all internet traffic according to recent report\cite{cisco-report}.  If large scale video flows stream the internet without any congestion control mechanism, the bandiwdth compitition would lead to packet drops and make the Internet congestion. This unresponsive behavior is unfair to self controled flows and wastes network resource. Even through  There were some works\cite{widmer2001survey}\cite{padhye1999model} making an effort to exploit congestion control scheme for UDP streaming media, none of these algorithms have application  in practice.

In an early stage, the implementation of congestion control on application layer for video streaming is quite scarce, due to the consideration that an insufferable QoE of VoIP connection would make the users give up video call, 
which can be seen as another mechanism of congestion avoidance. The network condition has changed in better direction and the 4K-UHD video is prepared its way to streame through network. As the popular of mobile devices, making video calls through IP network becomes quite common, and RTP-based media traffic has increased a lot.

To develop new congestion control algorithm for real time traffic has gained renewed attention in recent years, especailly since the open source of Web Real-Time Communication(WebRTC) which aims the interoperability of real time communication between browsers. It is used by many companies to develop video 
related application.
As pointed by\cite{ietf-rmcat-require}, all the flows transporting data across internet should implement congestion control scheme for internet congestion avoidance and promote fair bandwidth occupation. 
The IETF has initiated The RTP Media Congestion Avoidance Techniques (RMCAT) Working Group to develop congestion standards for interactive real-time media. 
And there are mainly three congestion control drafts under this working group, 
namely, GCC\cite{lundin2012google}, NADA\cite{zhu2013nada}, SCReAM\cite{rfc8298}.

One way to test the performance of these algorithms  is to compile the WebRTC source code and run it on the testbed topology recommended by\cite{sarker2015test}, just as the experiment done in\cite{carlucci2016analysis}. 
But in a practical test environment, the configuration of different network situations is restricted. 
And the complicated code structure of WebRTC makes its not a trivial task to implement new algorithm based on its code.

We work our way to get these algorithms running in ns-3\cite{ns3} and make a full comparison in term of protocol fairness, link queue delay and protocol competence of all three RMCAT algorithms performance. 
The simulation code of NADA\footnote{https://github.com/cisco/ns3-rmcat} on ns3 was already released by its author. So the main work was mianly focused on the implementation of GCC and SCReAM. And the simulation code of this work can be downloaded\footnote{https://github.com/sonyangchang/rmcat-ns3}.
And some drawbacks of the simulation experiments should be put forward here, a perfect encoder is assumed whose data output rate can be adjusted immediately and the encoder delay is not considered too.

We may have the bold to claim that we are the first to make a thorough analysis of all three RTP-base congestion control algorithms in a simulated environment.

The rest of this paper is orgianised as follows. Section 2 provides a brief review of related work on congetion control. Section 3 describes the  algorithms involved in experiments in detail. In Section 4, the simulation reuslts are presented and analysed. Section 5 is the conclusion.
\section{Related Work}

The congestion control is the most important part of network and  its main goal is to prevent network collapse and guarantee a reasonably fair bandwidth allocation among network users. 
Why does the AIMD congestion control law proposed by Jocobson becomes a guideline? 
And there have been thousands of papers pivoted around it since its birth thirty years ago. 
It solves a centralized network bandwidth allocation problem in a distributed way with limited feedback information 
without implanting a global network operator, 
which of course is infeasible considering the large scale of the internet. 
Jocobson’s method was later theorized by Kelly through introducing the Utility maximization function in\cite{kelly1998rate}. 
The distributed bandwidth allocation through implementing congestion control algorithm on end users, 
the limited feedback information for end user to adapt its sending rate, 
the heterogeneity of network, e.g. the wire network and the wireless, and the evolution of internet techniques, 
are the main reasons for continuous emergence of tcp congestion variants. 

Most congestion control algorithms are based on packet loss signal, e.g. Reno, Bic, Cubic. 
Due to the excessive buffer in current router equipment, 
the loss based algorithms tend to fill queue buffer full and cause high latency, 
which is notorious for buffer bloat\cite{gettys2011bufferbloat}. 
The additive increase process of congestion window makes the network resource utilization quite low, when there is more bandwidth available
The drawbacks of loss based algorithms encourge researchers to work out better soulutions. 

The idea of taking delay as network congestion signal has been proposed as early as 1989\cite{jain1989delay}. 
There are other delay based congestion control algorithms developed later 
such as TCP Vegas\cite{brakmo1995tcp} and TCP FAST\cite{wei2006fast}.
Those algorithms actively increse its congestion window when the delay remains small, and reduce its congestion window when the delay exceeds certain threshold to let the network to drain the building up queue. 
The delay based algorithms can infer link congestion earlier. When sharing bottleneck with loss based flows, delay based flows tend to get starvation. 
This drawback blocks its application in practice. 

LEDBAT\cite{rossi2010ledbat} and TCP LP\cite{kuzmanovic2003tcp} use one way delay to infer congestion. 
These two algorithms will increase congestion window to efficiently exploit network capacity if bandwidth is available 
and yield bandwidth to more urgent TCP flows. 
This feature makes them quite fit for low priority file transfer for example bit-torrent file sharing. 
CDG\cite{hayes2011revisiting} was crafted to employ RTT gradient to infer congestion with the goal of coexisting with loss based algorithms. 
TCP Westwood\cite{mascolo2001tcp} controls its sending rate by measurement the available bandwidth via 
the return rate of packet acknowledgement. 
Verus\cite{zaki2015adaptive} was designed for cellular networks 
considering the channel bandwidth fluctuations in radio links. 

The most recently remarkable work was TCP BBR\cite{cardwell2016bbr}, which 
claimed itself as congestion based congestion control algorithm by actively 
probing available bandwidth to converge the optimal point, namely, 
maximizing packet delivery rate while minimizing delay and loss.

Real time streaming applications often send data over UDP due to the strigent time requirement. The congestion avoidance phase is mostly carried out by regulating the data generating rate such as TFRC\cite{padhye1999model}. The rate control of TFRC was based on  the TCP throuphput formula deduced in\cite{padhye1998modeling}, which enables a stable rate for end user. The low latency requirement was taken into consideration by newly proposed algorithms GCC, NADA, SCReAM. GCC takes one way delay gradient for congestion signal, NADA takes an aggregated delay signal and SCReAM takes one way delay for rate control.  Rebera\cite{kurdoglu2016real} proactively measures the available bandwidth by packet trains, and determines a rate budget predicted from history    information for the video encoder. It aims to provide easy rate adaption, low eoncoding complexity and low delay.

The congestion control goal and requirement for real-time interactive media are described in\cite{ietf-rmcat-require}. The streaming data should be generated at a rate as close the available bandwidth as possible while keeping a low end to end delay. The media flow should get a fair share of bandwidth and do not starve or get starvation when coexisting with other type flows.

According to above description, the criteria for evaluation of the real time traffic 
congestion control algorithms can thus be obtained: low link queue occupation for low latency, 
reasonable bandwidth share in the presence of TCP flows, 
responsiveness in consideration of link bandwidth fluctuation. 
And we compare three algorithms performance in these terms.

\section{algorithm description}
This part briefly describes the algorithms involved in our experiments.
The GCC algorithm exploits one-way delay gradient as control signal. 
The old version of GCC has two components: a delay based congestion controller, 
running at the receiver side, computes a rate $A_r$ according to frame delay which is fed back through RTCP Receiver Estimated Maximum Bitrate (REMB) report; 
a loss based controller running at sender side, computes a target bitrate $A_s$ which shall not exceed $A_r$. 
Kalman filterer is adopted at the receiver side to compute the link queue delay gradient. 
In newer version of WebRTC, the congestion control logics have all been moved to the sender side. 
A trend-line filter has been introduced for congestion inference. 
We refer here the old version WebRTC congestion control based on kalman filter as REMB-GCC 
and the newer version based on trend-line filter is referred as TFB-GCC (transport feedback GCC). 
The algorithm designers have published several papers on REMB-GCC, 
please refer to\cite{carlucci2016analysis} for more information. 
We will describe the TFB-GCC in detail considering there is no published paper 
to recount its working principle.
\begin{figure}
\includegraphics[height=3in, width=3in]{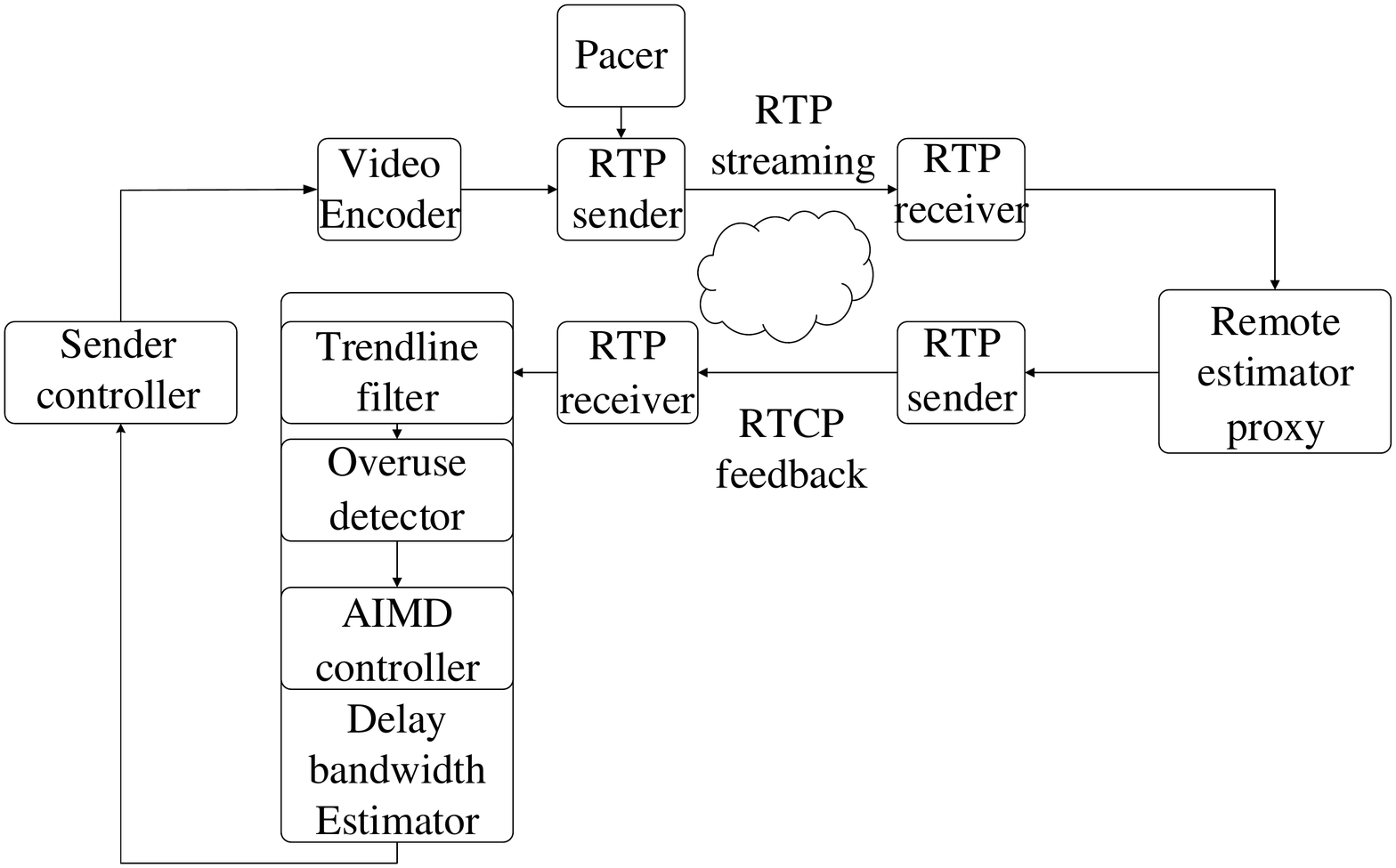}
\caption{The TFB-GCC architecture}
\label{Fig:WebRTC-tfb}
\end{figure}

\begin{figure}
\includegraphics[height=3in, width=3in]{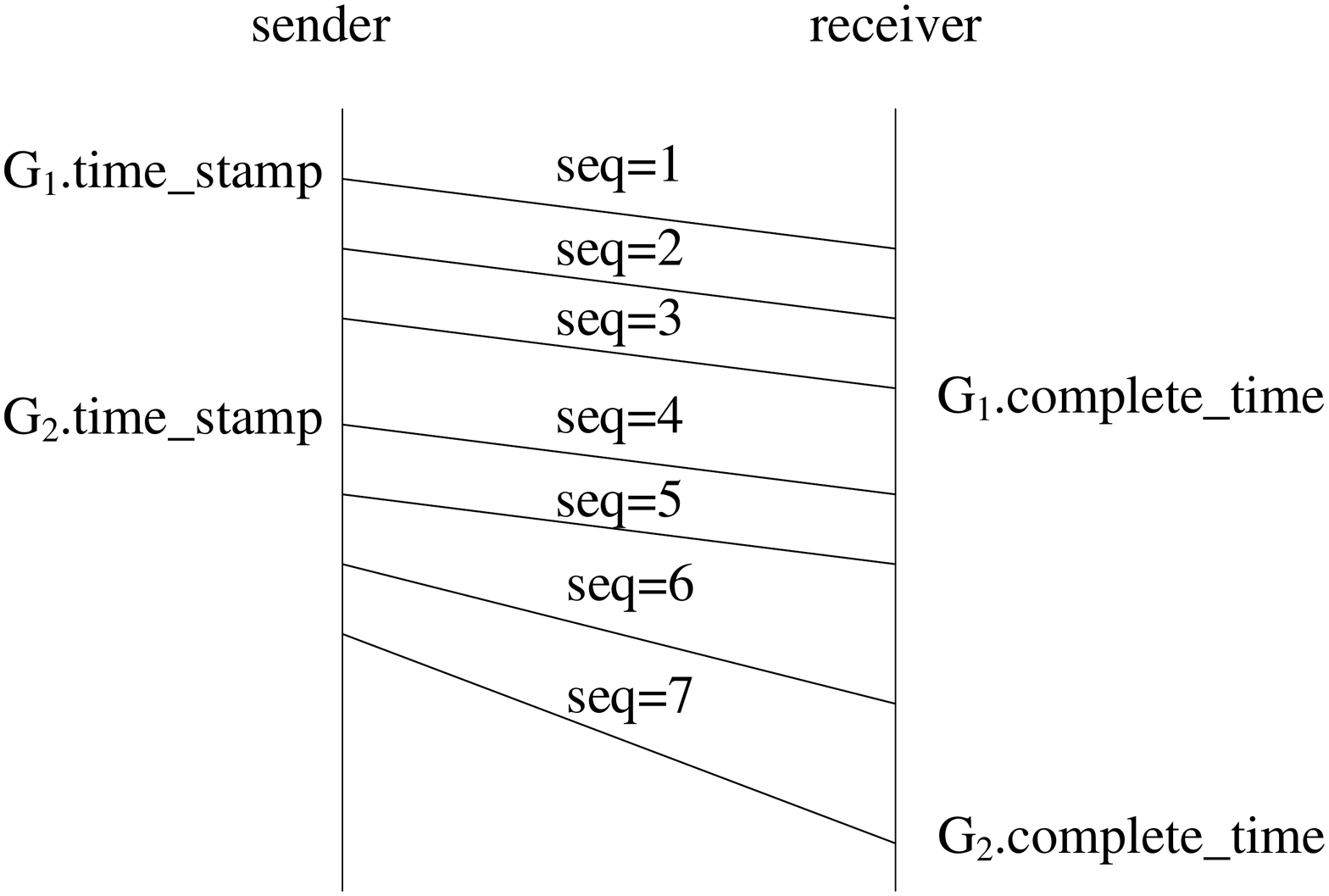}
\caption{Packets groups}
\label{Fig:trendline}
\end{figure}
The delay bandwidth estimator which originally functions at the receiver side 
has been moved to the logic on the sender as shown by Figure ~\ref{Fig:WebRTC-tfb}. 
So the receiver just feedbacks packets arrival time to the sender through 
the RTCP extensions for transport-wide congestion control\cite{holmer2015rtp}. 
The feedback message will be sent at an adaption interval according to bandwidth. 
When the feedback message arrives, the sender extracts out the sending out timestamp and the arriving time of a sent packet, 
and divides them into groups by length of five milliseconds as showed by Figure ~\ref{Fig:trendline}.

The packets group is similar to the frame notation in\cite{carlucci2016analysis}  for the purpose of channel overuse detection. 
The time\_stamp is the time sending out the first packet and complete\_time is the time of last packet arriving to the destination of the same group. 
The i-th group one-way delay variant is computed as follows:
\begin{equation}
\begin{aligned}
d_{i}={(G_{i}.complete\_time- G_{i-1}.complete\_time)}\\
{-(G_{i}.timestamp- G_{i-1}.timestamp)}.
\end{aligned}
\end{equation}

Then compute the accumulated delay:
\begin{equation}
acc\_delay_{i}=\sum_{j=1}^{i}delta\_ms_{j}.
\end{equation}

And then smooth the delay signal with a coefficient alpha by default 0.9.
\begin{equation}
\begin{aligned}
smoothed\_delay_{i}=smoothing\_coef*smoothed\_delay_{i-1}\\
+(1 - smoothing\_coef)*accu\_delay_{i}
\end{aligned}
\end{equation}

A linear regression was carried out in trend-line filter with input values of(x,y).
\begin{equation}
\begin{aligned}
(x,y)\Rightarrow(G_{i}.complete\_time- G_{1}.complete\_time\\
,smoothed\_delay_{i}).
\end{aligned}
\end{equation}
\begin{equation}
\begin{aligned}
trendline\_slope=\frac{\sum\nolimits(x_i-x_{avg})(y_i-y_{avg}) }{ \sum\nolimits(x_i-x_{avg})^2}
\end{aligned}
\end{equation}

The trendline slope is a reflection of link queue status. When the link queue length increases, the packets inter-arriving space tends to increase also.
The overuse detector compares the value of trendline slope with a dynamic threshold to 
decide if the channel is in the state of underuse or overuse. 
The dynamic threshold is explained by the designer\cite{carlucci2017congestion} to tune the sensitivity 
of the algorithm. A small threshold will make the detector quick detect the channel state changes 
but with the drawback of overreacting in case of noise. 
A large threshold would make the algorithm robust to noise 
but sluggish to channel state change. And a constant threshold would make 
the GCC flows starvation in competing with loss based TCP flows as reported by\cite{de2013understanding}.
After the overuse detector computes out the channel state, 
the AIMD controller adjusts the bitrate accrording to the equation:
\begin{equation}
\begin{aligned}
A(t_i)=\left\{
\begin{array}{rcl}
A(t_{i-1})+\overline A & & {Increase}\\
\beta R(t_{i-1}) & & {Decrease,}\\
A(t_i) & & {Hold.}
\end{array} \right.
\end{aligned}
\end{equation}
where $\beta=0.85$, and $R(t_{i-1})$ is the average receiving rate estimated at the sender side 
based on feedback message. The value of $\overline A$ is depended on the rate control region. 
After the rate is decreased, the controller would set the rate control region in state of 
near-max. After the channel is detected underuse and the control region in near-max state, 
the AIMD controler would additively increase rate, 
otherwise, the rate is multiplicatively increased.  

There is a detailed description and comparison between GCC and NADA on the WebRTC codebase 
platform in\cite{carlucci2016congestion}. And we shall not rehear here.

\begin{figure}
\includegraphics[height=3in, width=3in]{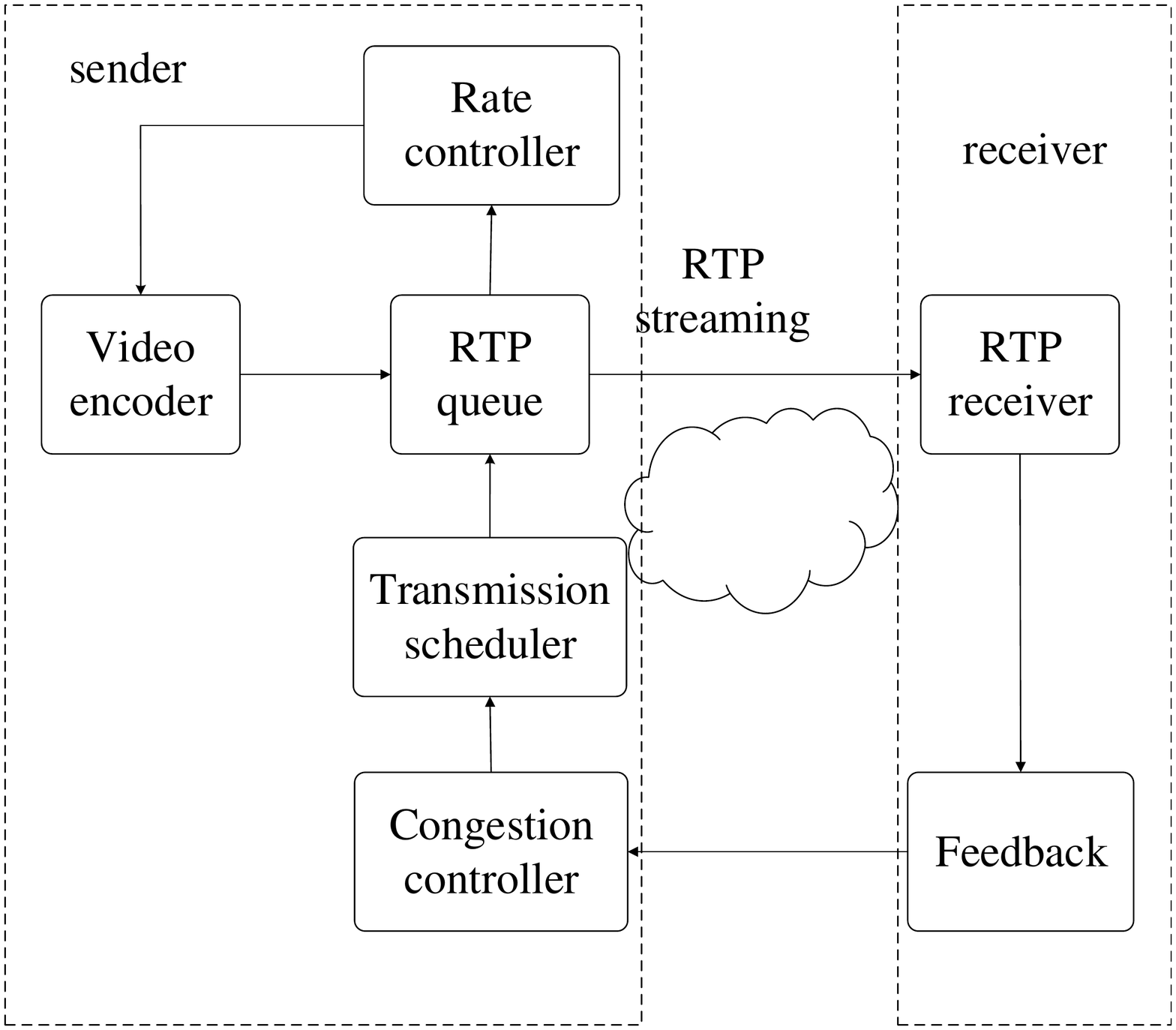}
\caption{The rate adaption arthitecture of SCReAM }
\label{Fig:scream}
\end{figure}
SCReAM basically controls the upper limit packets in flight by sliding congestion window.
The rate control architeture is showned in Figure ~\ref{Fig:scream}. 
The receiver will feedback the timestamp of received packet with the highest sequence number and 
an acknowledgement vector to indicate the reception or loss of previous packets. 
It’s congestion controll method was inspired by LEDBAT, 
which has claimed for the low network queue delay purpose by inferring congeestion earlier. It's congestion controller  adjusting the congestion window by Algorithm~\ref{alg:one}. 
When the one-way queue delay under the target queue delay, 
the algorithm will increase the congestion window, otherwise decrease the window.

The rate controller of SCReAM adjusts the encoder rate based on RTP queue delay, 
packet transmitting rate and acknowledge rate.

\makeatletter  
\def\BState{\State\hskip-\ALG@thistlm}  
\makeatother
\begin{algorithm}  
\caption{SCReAM Window Control Algorithm}\label{alg:one}  
\begin{algorithmic}[1]  
\Procedure{Incoming feedback}{}\\  
$ackedOwd\gets feedback.timestamp-$\\
$transmitted.timestamp;$\\
$baseOwd\gets min(baseOwd, ackedOwd)$;\\
$queueDelay\gets ackedOwd- baseOwd$;\\
$offTarget\gets (queueDelayTarget-$\\
$queueDelay)/queueDelayTarget$;\\  
\If{$offTarget>0$}\\
$cwnd\gets cwnd+gainUp*offTarget*bytesNewlyAckedLimited$
$*mss/cwnd$;
\Else \\
$cwnd\gets cwnd+gainDown*offTarget * $\\
$bytesNewlyAcked* mss / cwnd$;\\ 
\EndIf 
\EndProcedure  
\end{algorithmic}  
\end{algorithm}  
\section{Simulation comparison}
\begin{table}  
\caption{Network configuration}
\label{tab:configuration}  
\begin{tabular*}{8cm}{lll}  
\hline  
Bandwidth & Path Tansmission Delay& Path Queue Buffer \\  
\hline  
2Mbps  &100ms& 300ms*2Mbps\\ 
\hline  
\end{tabular*}  
\end{table}
A point to point topology as suggested by\cite{sarker2015test} was created on ns3 environment 
with link configuration in Table \ref{tab:configuration}. 
All the experiments were running about 200 seconds.
\subsection{Protocol Responsiveness}
Considering the popular of mobile device, the RTP-based media over mobile phone is quite common. 
The cellular  access  can present drastic change in channel bandwidth in a span of short time. The rate adaption algorithm for conservational video over wireless links should be reacted quickly to network change and operates in a wide range of bandwidth. When the link bandwidth decreases, the video generator keeping the rate before 
would make the link queue build up and the end latency increase, and can not make fully use of bandiwth resource when link bandwidth increases. 

In experiment, the link bandwidth is changed every 20 seconds from 500kbps to 2Mbps. The link is exclusively occupied by single GCC, NADA, or SCReAM flow.
\begin{figure}
\includegraphics[height=2.5in, width=3in]{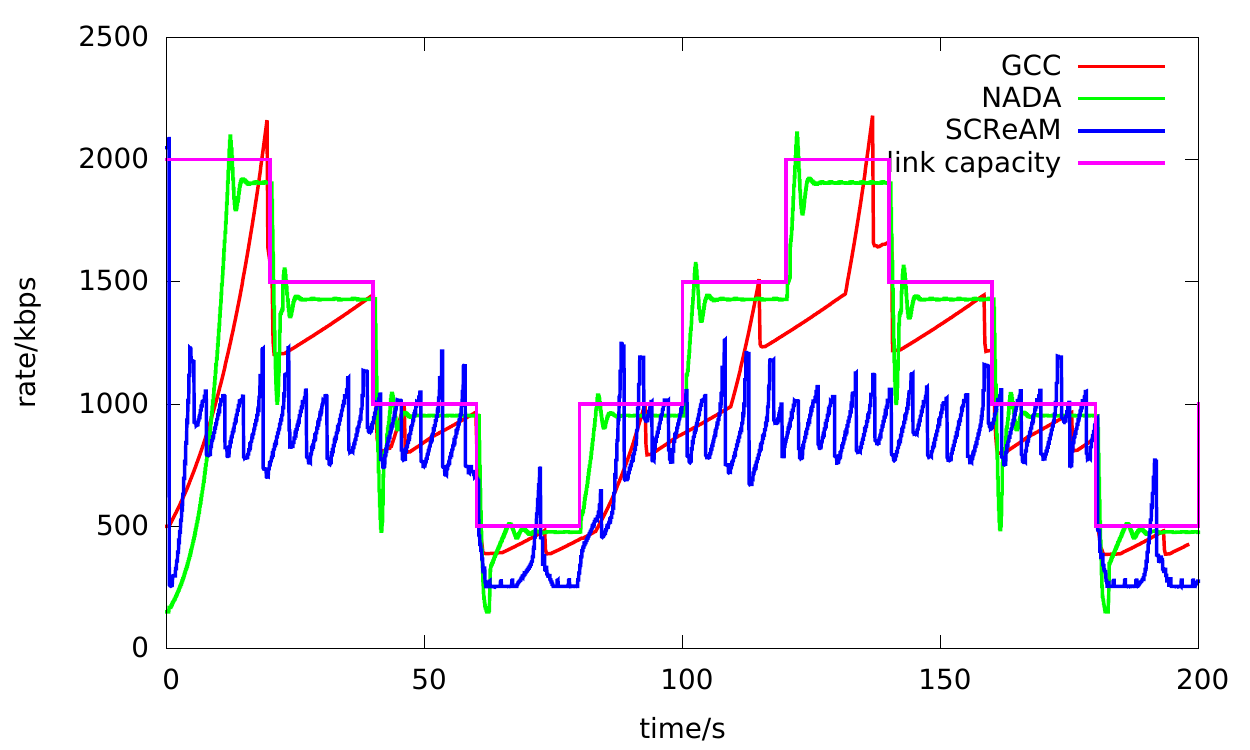}
\caption{The responsiveness of RMCAT protocol}
\label{Fig:responce}
\end{figure}
\begin{figure}
\includegraphics[height=2.5in, width=3in]{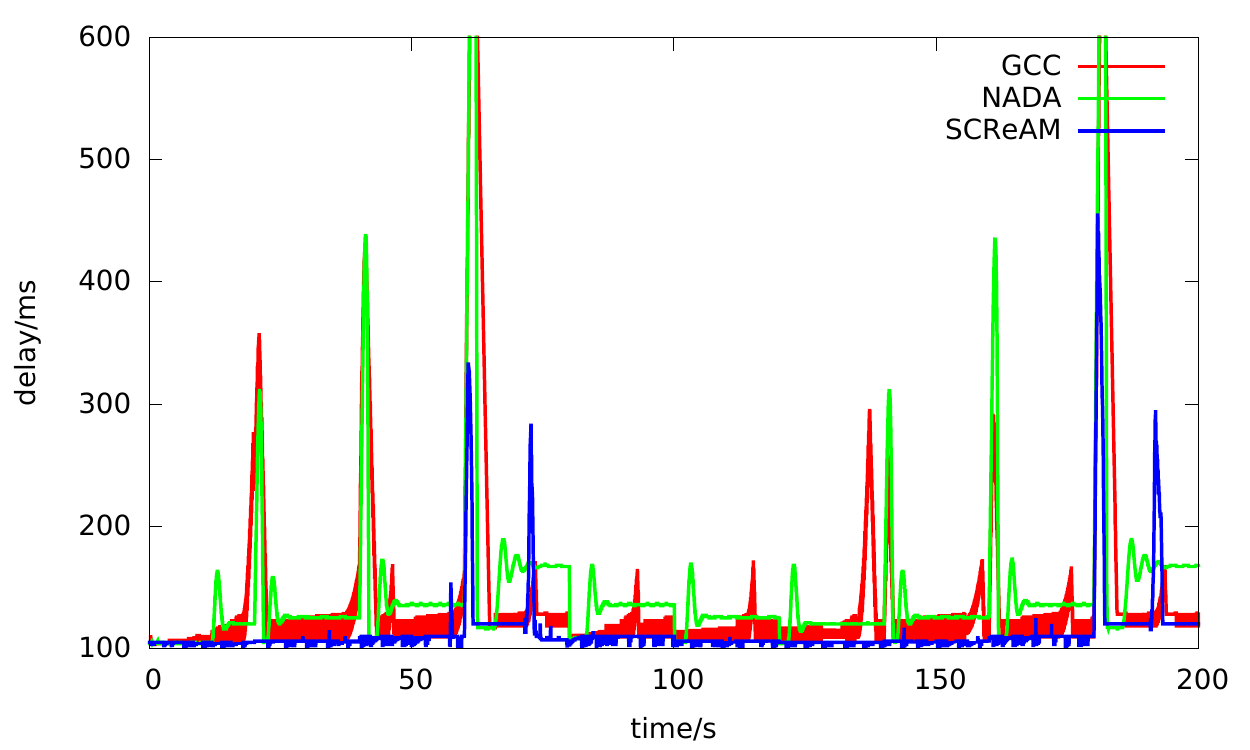}
\caption{Packet one way delay}
\label{Fig:delay}
\end{figure}
\begin{table} 
\caption{Average link utilization}
\label{tab:bw_utilization} 
\scalebox{0.7}{
\begin{tabular}{|l|ccccc|}
\hline
\diagbox{Protocol}{Utilization}{Time(s)} & 0-20 & 20-40 & 40-60& 60-80& 80-100\\
\hline
GCC & 56.7881\%&88.0983\%&89.2792\%&86.1947\%&71.5772\%\\
NADA &80.4109\%&95.5376\%&95.7954\%&98.6866\%&92.6458\%\\
SCReAM&43.41\%&61.0819\%&87.5441\%&62.7871\%&76.7606\%\\
\hline
\end{tabular}}
\end{table} 

The results in Figure~\ref{Fig:responce} have clearly shown the reaction difference of these protocols 
when link capacity changes. The AIMD controller of GCC for rate adjustment 
is the reason of its sawtooh rate curve, which makes its slow convergence rate . 
In comparison, NADA can quickly respond to network change and stabilizes its encoder 
rate in proximity to link capacity. SCReAM is sensitive to capacity decrease, but reacts sluggishly to capacity increase.

The average link bandwidth utilization is measured in Tabel~\ref{tab:bw_utilization}. NADA has the highest channel utilization and SCReAM makes the lowest channel utilization which may cause by its rate ramp-up parameter

From the one-way delay variation curve in Figure~\ref{Fig:delay}, SCReAM reaches its claimed goal by having 
the lowest queue delay occupation close to  one-way link transmission delay. 
NADA and GCC make link queue build up to some extent. All three protocols show instantaneous delay spike when faced sharp bandwidth decrease. 

\subsection{Intra Protocol Fairness}
Protocol fairness is an important indication to reflect whether an end user converges to 
a fair bandwidth share with other flows passing through the same link. 
In this experiment, three flows exploiting the same congestion control protocol were 
initiated at different time point over a bottleneck link. 
The second flow started after 40s later of the first flow and the third flow started at 80s. 
The link capacity keeps to be a constant value 2Mbps during the simulation. 

\begin{figure}
\includegraphics[height=2.5in, width=3in]{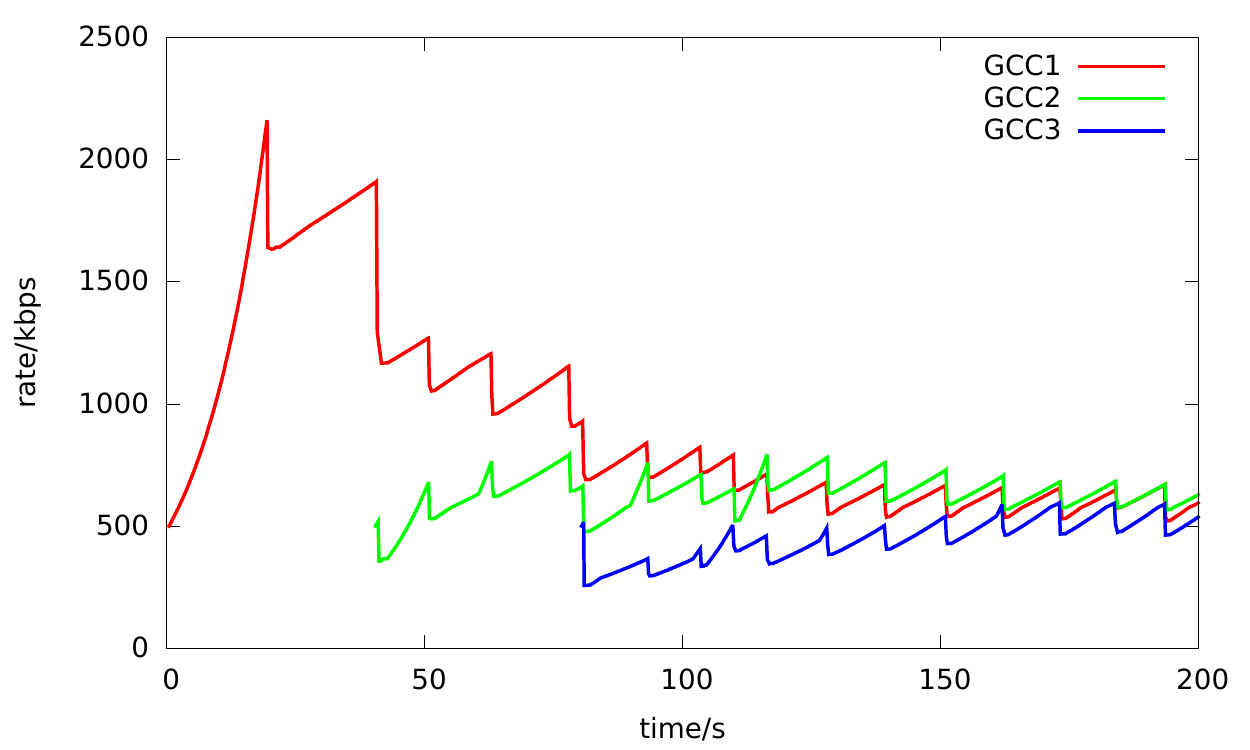}
\caption{Sending rate of GCC flows}
\label{Fig:webrtc3}
\end{figure}

\begin{figure}
\includegraphics[height=2.5in, width=3in]{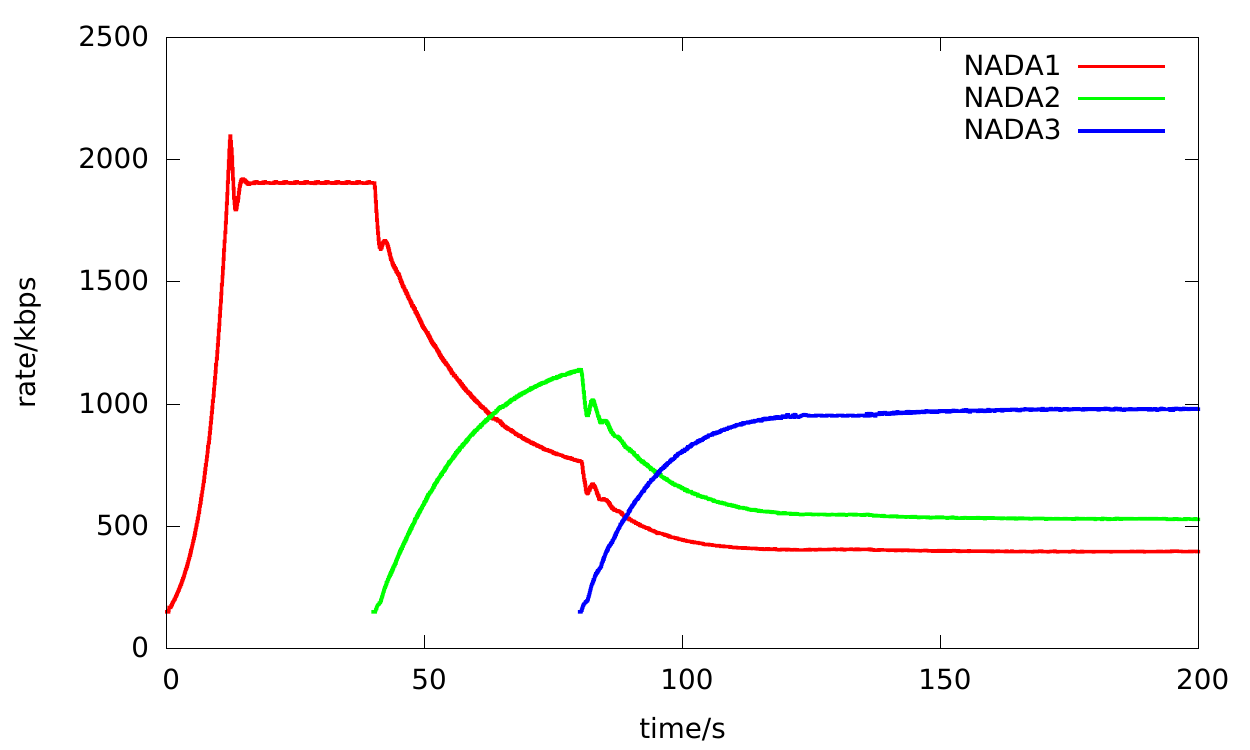}
\caption{Sending rate of NADA flows}
\label{Fig:nada3}
\end{figure}

\begin{figure}
\includegraphics[height=2.5in, width=3in]{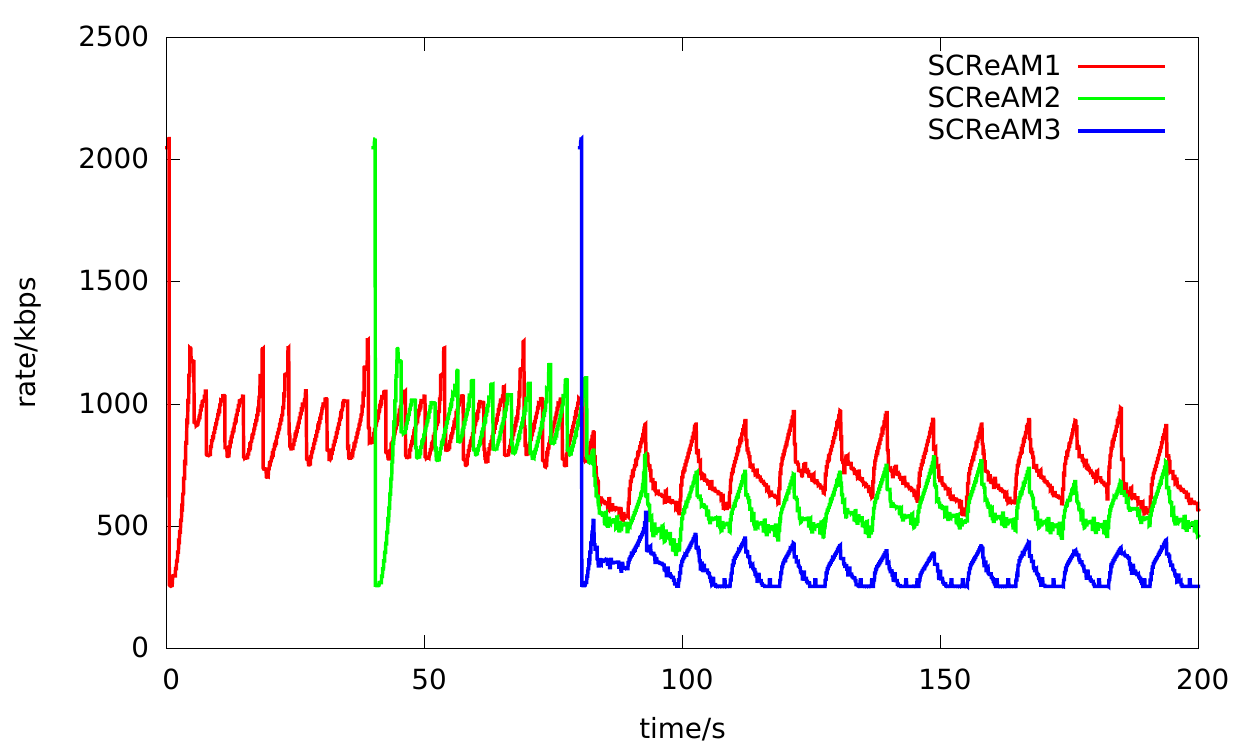}
\caption{Sending rate of SCReAM flows}
\label{Fig:scream3}
\end{figure}

In Figure ~\ref{Fig:webrtc3}, the rates of all three GCC flows after 150 seconds are very close, 
indicating the GCC protocol has fine fairness property. 
It’s worth noticing the NADA protocol suffers from “late-comer effect” in Figure ~\ref{Fig:nada3}, 
the later coming flow data sending rate is higher that the flows initiated before. This result is different from the conclusion in\cite{carlucci2016analysis}. The "late-comer effect" may be caused by its delay value function updated in new version of NADA. The SCReAM protocol in Figure ~\ref{Fig:scream3} shows no sign that the flows converge to a fairness rate. Due to the effect of link queue building up, the rate adjustment of SCReAM shows oscillation.
\subsection{Protocol Competence}
\begin{figure}
\includegraphics[height=2.5in, width=3in]{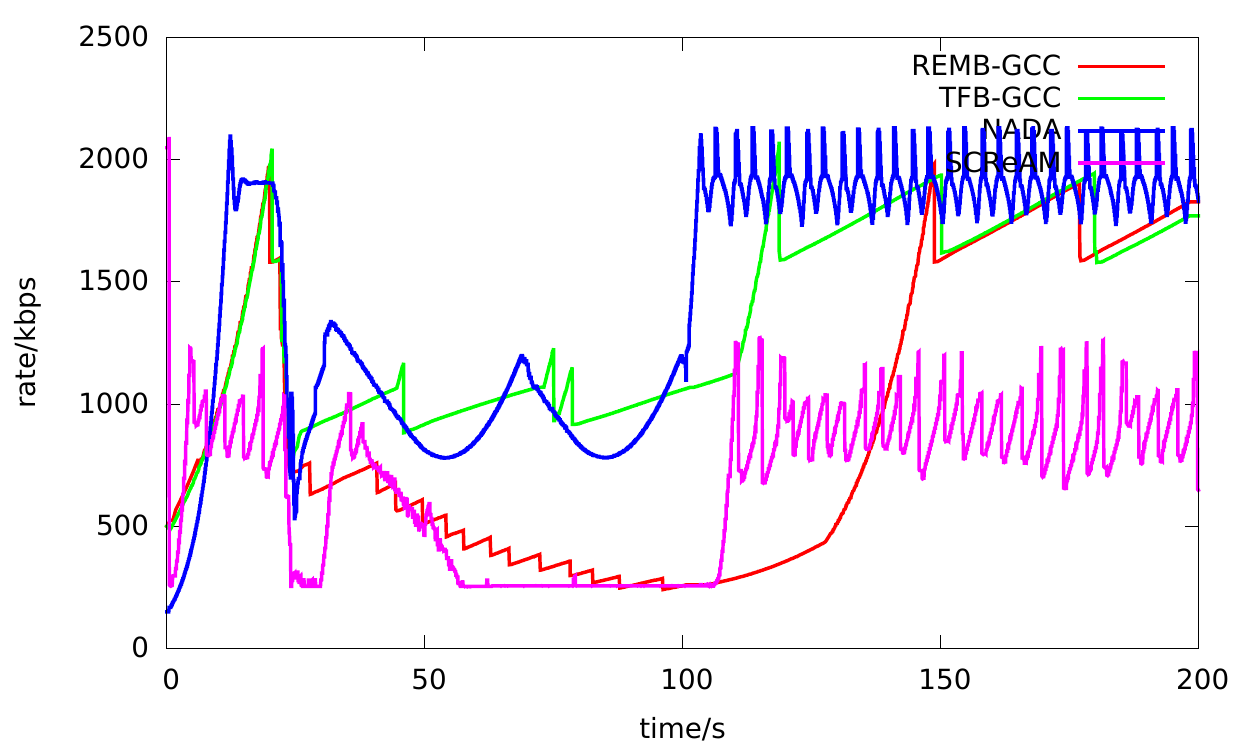}
\caption{RMCAT flow sharing links with tcp}
\label{Fig:rmcat-tcp}
\end{figure}
In real network, a network routing can be shared by many flows, 
which may exploit different congestion control protocol. 
In facing of background traffic, the ability to make a reasonable bandwidth 
occupation of a protocol is quite important.
For testing purpose, an experiment was designed for a RMCAT flow sharing link 
with a TCP Reno flow. The TCP flow was started at 20 seconds and stopped at 100 seconds. 
Even through the REMB-GCC was deprecated in new version of WebRTC, 
we test its performance here. The results is shown in Figure~\ref{Fig:rmcat-tcp}.

When the TCP data flows into the link, the REMB-GCC keeps yielding its bandwidth 
until reaching the smallest point. TFB-GCC and NADA can strive for a reasonable sending rate. 
SCReAM also decreases its rate to the minimal default rate due to the TCP flow occupying 
much link queue resource. When the link queue link on the merge of full, 
packet loss event would happen and the TCP flows would half it congestion window 
to relieve the link from congestion, the queue delay decrease signal would make NADA and TFB-GCC increase its rate. This explains why the rate curve of NADA and TFB-GCC have increase tendency process 
during the presence of TCP flow. 
When TCP flow exits off the network at the time point 100, NADA can make faster increase to reach a rate near the link capacity than TFB-GCC. 
It should be pointed out NADA flow shows oscillation even when the tcp flow withdraws from the link. This may cause by packet loss during the competence period in the presence of tcp flow. Pakcer loss makes the aggregated  congestion signal increase and introduces a penalty to NADA rate.
\subsection{Packet loss resistance}
\begin{figure}
\includegraphics[height=2.5in, width=3in]{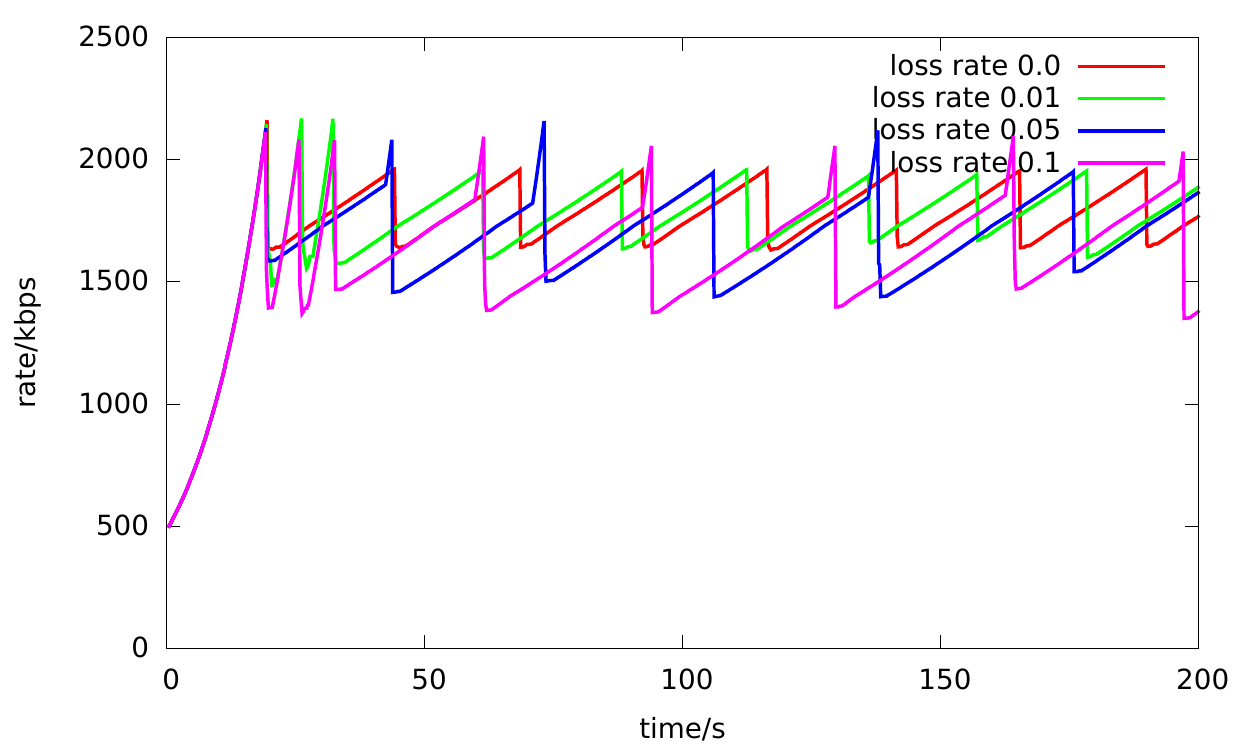}
\caption{GCC rate change in random packet loss link}
\label{Fig:webrtc_loss}
\end{figure}
\begin{figure}
\includegraphics[height=2.5in, width=3in]{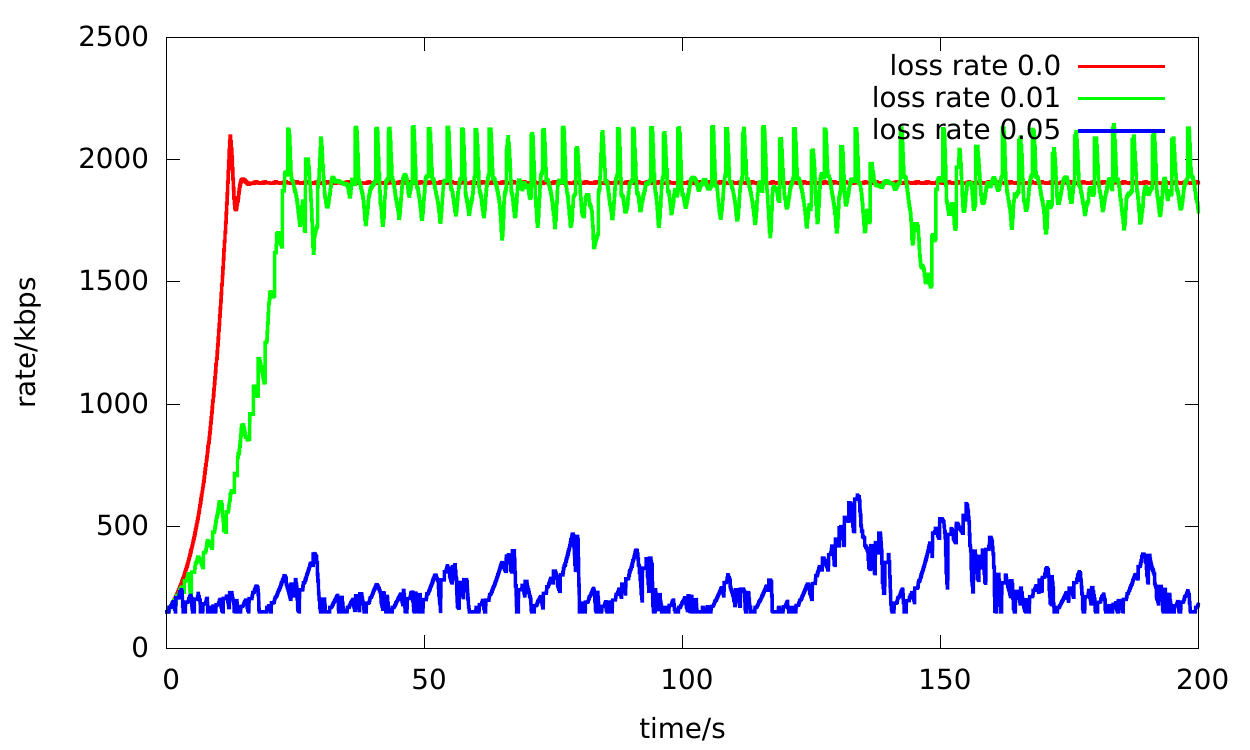}
\caption{NADA rate change in random packet loss link}
\label{Fig:nada_loss}
\end{figure}
\begin{figure}
\includegraphics[height=2.5in, width=3in]{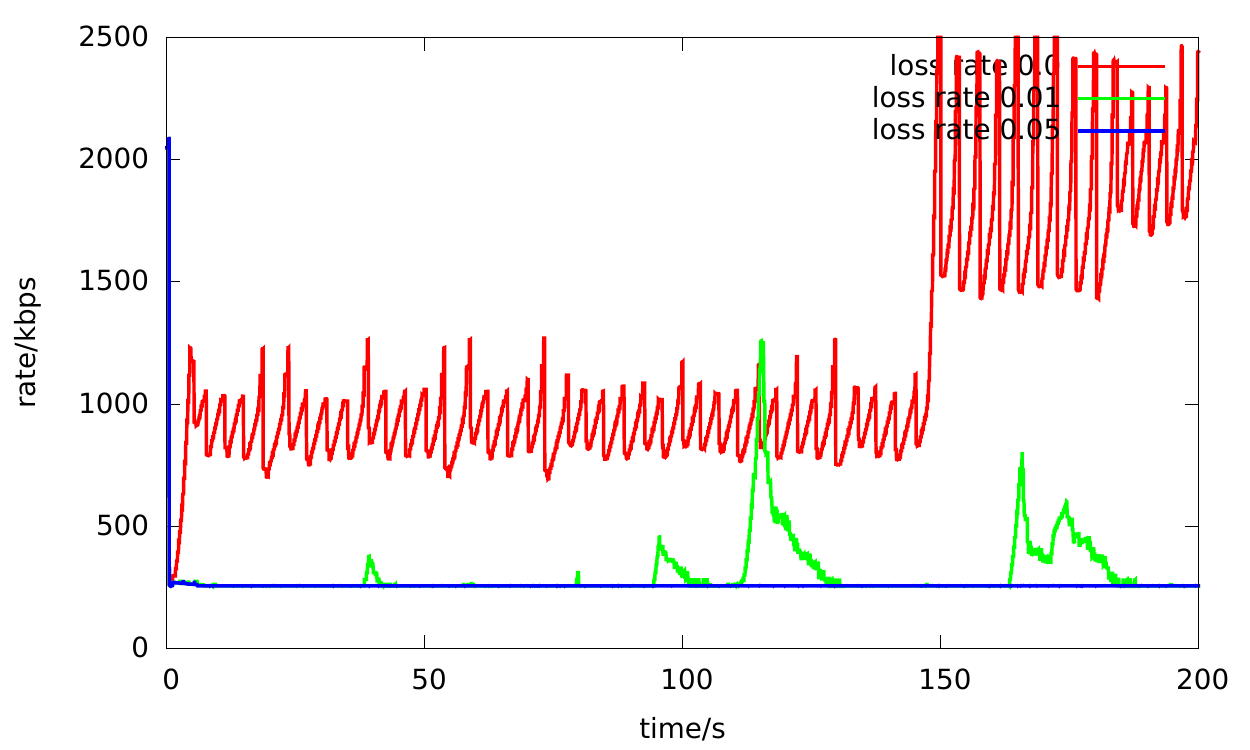}
\caption{SCReAM rate change in random packet loss link}
\label{Fig:scream_loss}
\end{figure}
\begin{table} 
\caption{Capacity utilization in lossy link}
\label{tab:loss_bw_utilization} 
\scalebox{0.7}{
\begin{tabular}{|l|ccc|}
\hline
\diagbox{Protocol}{Utilization}{loss rate} & 0.0\% & 1\% & 5\%\\
\hline
GCC &86.32\%&85.81\%&82.05\%\\
NADA &94.28\%&92.65\%&14.40\%\\
SCReAM&57.62\%&15.30\%&13.04\%\\
\hline
\end{tabular}}
\end{table}
In wireless links, packet loss may cause by wireless link interference, channel contention and errors. A protocol takes random packet loss as congestion signal  and  reacts it by rate decreasing will  have degenerated performance and low channel utilization. In this experiment, the link is configured with different random packet loss rate, and the link is monopolized by a single flow during the simulation.

As Figure~\ref{Fig:webrtc_loss} shows, GCC flow is not quite appreciably affected by random packet loss. As the packet loss rate increases, NADA and SCReAM decrease bitrate quite obvious, shown respectively in Figure~\ref{Fig:nada_loss} and Figure~\ref{Fig:scream_loss}. In the case of 5\% packet loss, GCC can hold  82.05\% channel utilization on average, and both NADA and SCReAM have quite low link utilization shown in Table~\ref{tab:loss_bw_utilization}.
\section{CONCLUSION}
The main work of this paper is the importation of GCC and SCReAM on simulated environment, 
and makes a full comparison of the three RMCAT algorithms in respect of protocol 
responsiveness, intra protocol fairness, inter protocol competence and performance in lossy link.

The area of network congestion control has developed for nearly thirty years. 
The implementation of congestion control for real time traffic is not a new idea. 
With the popular and influence of WebRTC, this research area again obtains researchers' attention. 
And with the evolvement of network technology and requirement of new application, 
the old tree of congestion control research areas always springs new sprouts such as TCP BBR.There is no once and for all solution for congestion control to fit all network situation. 

The results from simualtion are summarized here. GCC work well in intra protocol fainess but has saw tooth feature in dynamic links. NADA can quickly stablize its rate in dynamic links and has the most efficient network capacity utilization when the link is not affected by random loss, but suffers from "late comer effect". SCReAM retains the link queue delay in a low level but has low channel utilizaion. GCC has better perfomance in lossy links, which makes it particularly suitable for wireless network. To design a protocol with advantages of these algorithms should be our future work. The importation of these algorithms on ns3 platform would make the performance comparison of newly designed algorithms with the three easier. This simulation results will provide some reference for protocol designers. 


\bibliographystyle{unsrt}
\bibliography{rmcat-simulation}

\end{document}